\documentclass[9pt,twocolumn,twoside]{osajnl}

\journal{ol} 

\setboolean{shortarticle}{true}

\usepackage{siunitx}
\usepackage{graphicx}
\usepackage[justification=justified]{caption}

\title{Gas-plasma based generation of broadband THz radiation with 640 mW average power}

\author[1,*]{Joachim Buldt}
\author[1]{Henning Stark}
\author[1]{Michael Mueller}
\author[1,2]{Christian Grebing}
\author[1]{César Jauregui}
\author[1,2,3]{Jens Limpert}

\affil[1]{Institute of Applied Physics, Abbe Center of Photonics, Friedrich-Schiller-University Jena, Albert-Einstein-Str. 15, 07745 Jena, Germany}
\affil[2]{Fraunhofer Institute for Applied Optics and Precision Engineering, Albert-Einstein-Str. 7, 07745 Jena, Germany}
\affil[3]{Helmholtz-Institute Jena, Fröbelstieg 3, 07743 Jena, Germany}

\affil[*]{Corresponding author: joachim.buldt@uni-jena.de}

 \doi{\url{https://doi.org/10.1364/OL.442374}}

\begin{abstract}
We present a high-power source of broadband terahertz radiation covering the whole THz spectral region ($ 0.1-\SI{30}{\tera\hertz}$). The two-color gas plasma generation process is driven by a state-of-the-art Ytterbium fiber chirped pulse amplification system based on coherent combination of 16 rod-type amplifiers. Prior to the THz generation, the pulses are spectrally broadened in a multi-pass-cell and compressed to $ \SI{37}{\femto\second} $ with a pulse-energy of $ \SI{1.3}{\milli\joule} $ at a repetition rate of $ \SI{500}{\kilo\hertz} $. A gas-jet scheme has been employed for the THz generation, increasing the efficiency of the process to 0.1\%. 
The air-biased-coherent-detection scheme is implemented to characterize the full bandwidth of the generated radiation.
A THz average power of $ \SI{640}{\milli\watt} $ is generated, which is the highest THz average power achieved to date. This makes this source suitable for a variety of applications, e.g. spectroscopy of strongly absorbing samples or driving nonlinear effects for the studies of material properties.
\end{abstract}

\setboolean{displaycopyright}{true}

\begin{document}

\maketitle

The interest in powerful sources of terahertz (THz) radiation has been constantly growing over recent years. Increasing the available THz power reduces the data acquisition time, increases the signal to noise ratio and opens the path for novel applications. For example, with high-power THz radiation, applications in fundamental science \cite{Nuss1991,Razzari2009}, industry, biology \cite{Weightman2012,Ajito2011} and medicine \cite{Yang2016} can be addressed.  

Driven by the need of the applications for more powerful THz sources, different concepts have been studied to generate high power THz radiation. 
One very promising concept is the frequency conversion of high-power, ultra-short laser pulses in the near- and mid-infrared spectral region. Here, the most powerful sources so far have reached $ \SI{50}{\milli\watt} $ \cite{Buldt2020}, $ \SI{66}{\milli\watt} $ \cite{Saraceno_66mW} and $ \SI{144}{\milli\watt} $ \cite{SLAC_144mW}. For the further scaling of the generated THz radiation, either the efficiency or the driving laser power has to be increased. 

The highest efficiencies reported to date have been achieved with optical rectification (OR) using cryogenically cooled lithium niobate with up to $ \SI{3.8}{\percent} $ \cite{Huang2013,Huang2015} or organic crystals at room temperature \cite{Vicario2014,Ruchert2012,Hauri2011} with up to $ \SI{3}{\percent} $ \cite{Vicario2014} . 
However, neither of these sources have been scaled up to an average power beyond $ \SI{20}{\milli\watt} $ yet. 
The highest average powers demonstrated in \cite{Saraceno_66mW} and \cite{SLAC_144mW} were achieved using $ GaP $ and $ LiNbO_3 $ crystals. However, in these realizations, the efficiency of the conversion process is in the range of $ 6\cdot 10^{-4} $. Absorption of the driving laser as well as the terahertz radiation heats up these crystals and represents a major challenge in further increasing the generated average power. 

\begin{figure*}[!htb]
\centering
\includegraphics[width=0.75\textwidth]{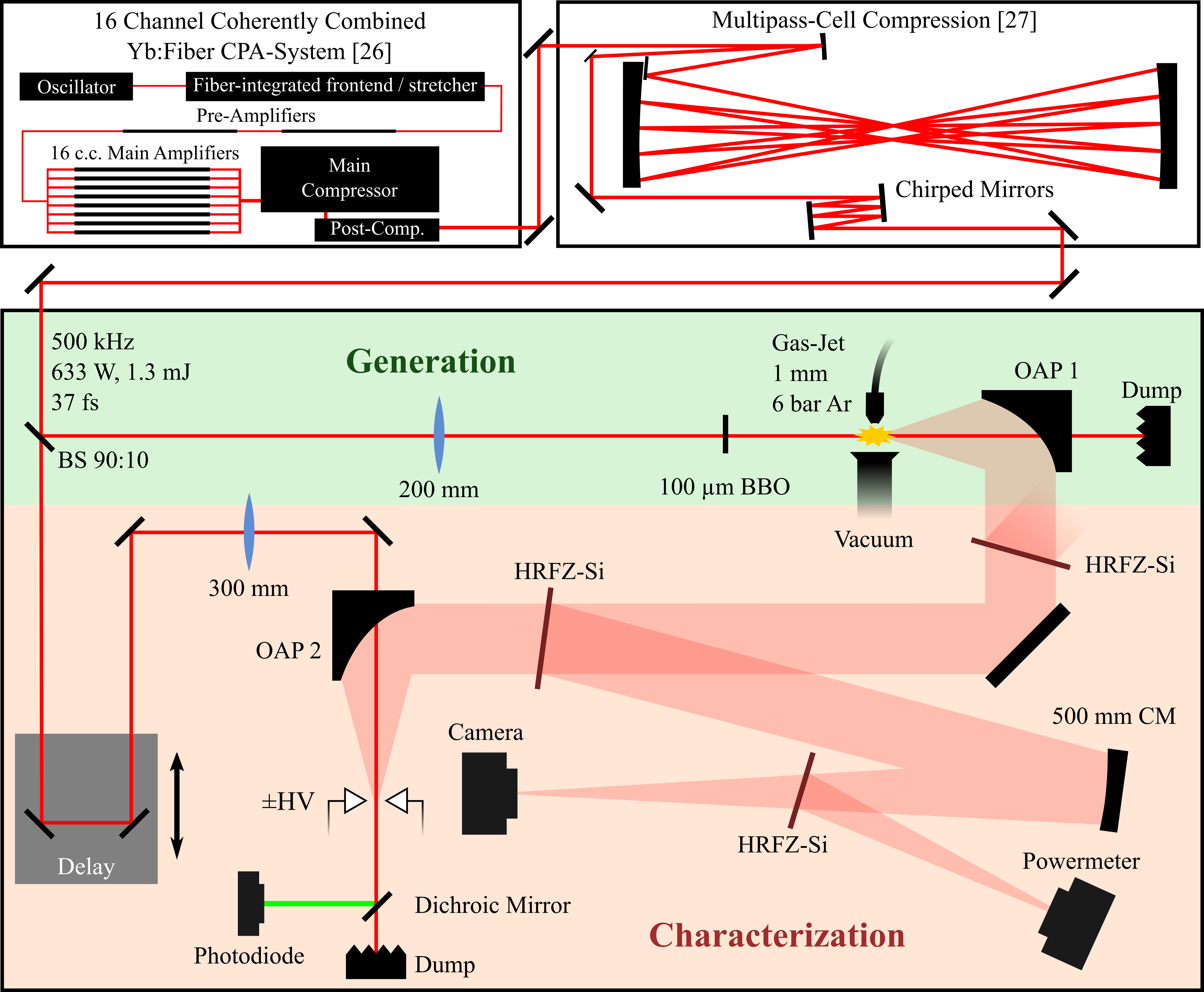}
\caption{Schematic of the setup used in this experiment. The pulses of the 16 channel coherently combined (c.c.) Yb:FCPA system are compressed from $ \SI{190}{\femto\second} $ to $ \SI{37}{\femto\second} $ pulse duration using a multipass cell. 
A beamsplitter reflects $ \SI{90}{\percent} $ of the pulse energy, which is used for the generation.
Here, the pulses are focused through a BBO crystal into a gas-jet. 
An off-axis parabolic mirror (OAP 1) collimates the THz radiation and the IR laser is dumped through a hole in the OAP. Residual laser light is blocked by a high-resistivity float-zone silicon filter (HRFZ-Si). For the simultaneous detection of the electric field, the beam profile and the average power of the THz radiation, additional HRFZ-Si wafer are inserted in the THz beam path to split the beam. The THz beam sample reflected by the second HRFZ-Si wafer for the power- and beam-profile measurement is focused with a 500 mm focal length curved mirror (CM). The electric field is measured with an ABCD \cite{ABCD} setup.}
\label{fig:setup}
\end{figure*}

\begin{figure}[!htb]
\centering
\includegraphics[width=\linewidth]{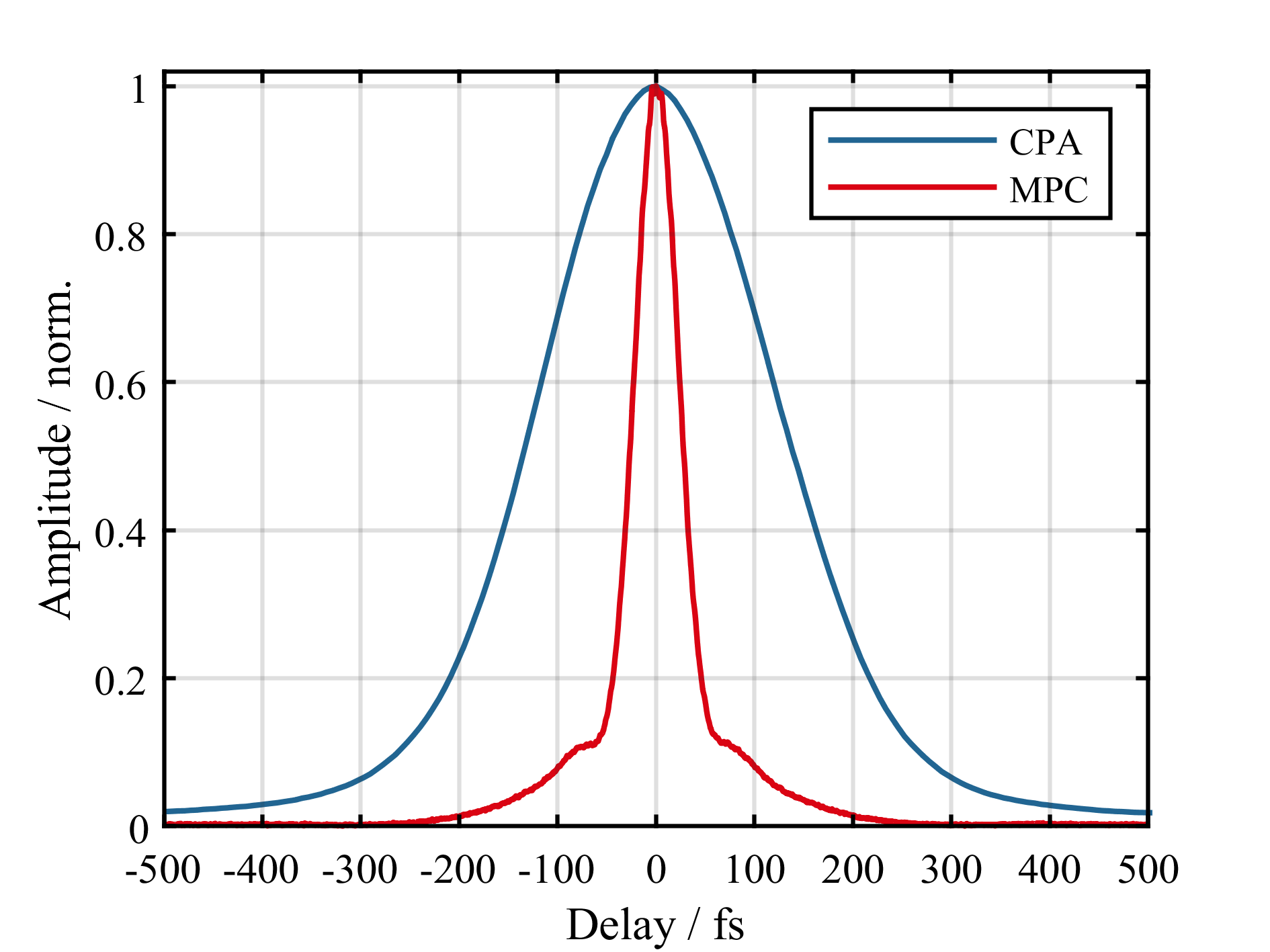}
\caption{Auto-correlation measurement of the pulses directly emitted by the CPA system (blue) and after the multipass-cell compression (red). The Yb:FCPA pulses have a duration of $ \SI{190}{\femto\second} $ and are compressed down to $ \SI{37}{\femto\second} $ in a multipass-cell compression (MPC) stage.}
\label{fig:ac}
\end{figure}

Another approach to THz generation is the two-color gas plasma scheme \cite{Cook2000}.
Here, the efficiency scales strongly with the driving wavelength \cite{WavelengthScaling0,WavelengthScaling1,WavelengthScaling2} which has allowed for a record efficiency of 2.36 \% with a $ \SI{3.9}{\micro\meter} $ driving wavelength \cite{Koulouklidis2020}. However, at these mid-infrared wavelengths the average power of the generated THz radiation is limited by the output power of the available lasers. Going to shorter wavelength allows for the direct use of highly developed ultrafast ytterbium lasers, which are able to deliver a power in excess of $ \SI{10}{\kilo\watt} $ \cite{Muller2020}. The efficiency of the generation process at this wavelength is still comparable to that achieved in $ GaP $ and $ LiNbO_3 $ \cite{Oh2014,Hah2017,Rodriguez2010}. In a previous publication \cite{Buldt2020} we demonstrated the generation of $ \SI{50}{\milli\watt} $ average power with an efficiency of $ 4\cdot 10^{-4} $. 

A major advantage of the gas plasma scheme is that the achievable bandwidth is limited only by the driving laser pulse duration and not by phase matching constraints as in crystal-based schemes.
For example, with sub-20 fs pulses a spectrum extending up to $ \SI{60}{\tera\hertz} $ has been demonstrated \cite{Piccoli2019}.
Another advantage of this gas-based approach is the circumvention of temperature and damage problems in the conversion medium as this can be constantly replenished. However, thermal effects in the plasma have been reported to cause a drop in the generation efficiency while increasing the repetition rate from $ \SI{6}{\hertz} $ to $ \SI{6}{\kilo\hertz} $ \cite{RepRateScaling}. In fact, no THz source based on gas-plasma with a repetition-rate over $ \SI{50}{\kilo\hertz} $ has been demonstrated so far and it has been an open question whether efficient THz generation in a gas at hundreds of kilohertz repetition rate is possible.

In this paper we demonstrate the generation of broadband single-cycle THz pulses at $ \SI{500}{\kilo\hertz} $ repetition rate with high efficiency, delivering a record-high average power of $ \SI{640}{\milli\watt} $.

The experimental setup is depicted in Fig. \ref{fig:setup}. The coherently combined, ytterbium-doped fiber chirped-pulse amplification (Yb:FCPA) system described in reference \cite{Stark2021} is used as the driving laser. The $ \SI{190}{\femto\second} $ long pulses centered at $ \SI{1030}{\nano\meter} $ and delivered by the Yb:FCPA system are further compressed in the multipass-cell compression (MPC) stage described in detail in reference \cite{Grebing2020}. The whole system is set to emit pulses at $ \SI{500}{\kilo\hertz} $ repetition rate with an average power of $ \SI{633}{\watt} $, corresponding to a pulse energy of $ \SI{1.3}{\milli\joule} $. An auto-correlation measurement of both the pulses delivered by the Yb:FCPA system and the pulses after the MPC stage are shown in Fig. \ref{fig:ac}. The fully compressed pulses have a pulse duration of $ \SI{37}{\femto\second} $. 

At the average power and pulse energy used in this experiment, the pulses cannot propagate in ambient air. Nonlinear absorption combined with the high average power creates a thermal lens \cite{Kartashov06}, which causes significant beam distortions that render the beam useless for any experiment. This effect can be avoided by enclosing the propagation of the fully compressed pulses in a noble-gas atmosphere. For that, in the CPA system \cite{Stark2021} a two-stage compressor design has been implemented, with a compact second compressor inside a vacuum-chamber purged with argon. Thus, the whole beam path of the fully compressed pulses is enclosed in vacuum-chambers filled with argon at different pressure. The post compressor and the beam path to the multipass-cell are kept at $ \SI{1}{\bar} $ argon. The multipass-cell itself is at $ \SI{250}{\milli\bar} $ argon and the subsequent beam path towards the THz experiment together with the experiment chamber are at $ \SI{150}{\milli\bar} $.

The compressed pulses coming from the multi-pass cell are split in two replicas, $ \SI{90}{\percent} $ of the pulse energy is used to drive the THz generation and $ \SI{10}{\percent} $ is used for air biased coherent detection (ABCD) \cite{ABCD}. The generation is based on the two-color gas plasma THz generation scheme \cite{Cook2000}. For that, the pulses are focused with a $ \SI{200}{\milli\meter} $ focal length lens through a $ \SI{100}{\micro\meter} $ thick BBO crystal to generate the second harmonic with around $ \SI{5}{\percent} $ conversion efficiency. Aligned with the focus is a gas-jet with $ \SI{1}{\milli\meter} $ diameter operated at a backing pressure of $ \SI{6}{\bar} $ argon. On the opposite side of the beam focus a vacuum tube is placed to constantly evacuate the gas. 
The localization of the gas-jet in the focus allows using a higher pressure in the generation region without the need for a chamber that has to withstand the mechanical load of the high pressure gas. At the same time, the constantly blown gas eliminates any thermal effects that might be caused due to the high driving laser power. It has been observed that thermal lensing in the gas can cause a drop in the efficiency of the generation process at repetition rates of up to $ \SI{6}{\kilo\hertz} $ \cite{RepRateScaling}. Unfortunately, no absolute values of efficiency are provided in \cite{RepRateScaling}, preventing the comparison of the efficiency at low and high repetition rates. In our experiment with the gas-jet no change in the efficiency has been observed when increasing the repetition rate from $ \SI{50}{\kilo\hertz} $ to $ \SI{500}{\kilo\hertz} $.

In the focus a plasma is created in which the THz radiation is generated. The THz radiation is collimated with a $ \SI{50}{\milli\meter} $ focal length off-axis parabolic mirror (OAP 1). The remaining driving laser radiation is dumped through a hole in the center of this OAP. Any further residual laser radiation is blocked by a high-resistivity float-zone silicon (HRFZ-Si) filter in the THz beam path.
For the optimization of the generation process, both the gas-jet and the focusing lens need to be adjusted precisely with respect to the collimating OAP. 


The THz radiation is focused with a $ \SI{100}{\milli\meter} $ focal length OAP (OAP 2) and superimposed with the $ \SI{10}{\percent} $ sample of the pulse which is transmitted through a hole in the OAP. Here a measurement based on the ABCD scheme \cite{ABCD} is used: Aligned with the focus are two electrodes connected to an alternating high-voltage of up to $ \SI{4}{\kilo\volt} $. The alternating voltage is synchronized with the pulse train emitted by the laser system and set to a frequency of $ \SI{50}{\kilo\hertz} $. In the focus the second harmonic of the laser pulse sample is generated. This is separated from the fundamental with a set of five dichroic mirrors, coupled into a fiber and detected with an avalanche photo-diode. With the help of a lock-in amplifier the second harmonic signal, synchronized with the alternating high voltage, is filtered. With a changing delay of the sample laser pulse, the relative temporal electric field of the THz pulse can be measured. 

An additional HRFZ-Si filter is inserted into the collimated THz beam for a further characterization of the pulse. This creates a sample which is then split in two parts to simultaneously measure the THz power (Gentec-EO THZ12D-3S-VP-D0 power meter), the beam profile (Swiss Terahertz, RIGI-L2 Camera) and the electric field for a full characterization. The real-time, full characterization allows also a precise alignment of the beam to achieve the maximum efficiency. The transmissivity and reflectivity of all HRFZ-Si filters have been calibrated for the THz radiation. This allows calculating back the generated radiation from the power measured in the sample. Note that the HRFZ-Si filters have a very flat spectral response and no spectral differences could be observed between the ABCD measurements with and without the filters.

\begin{figure}[!tb]
\centering
\includegraphics[width=\linewidth]{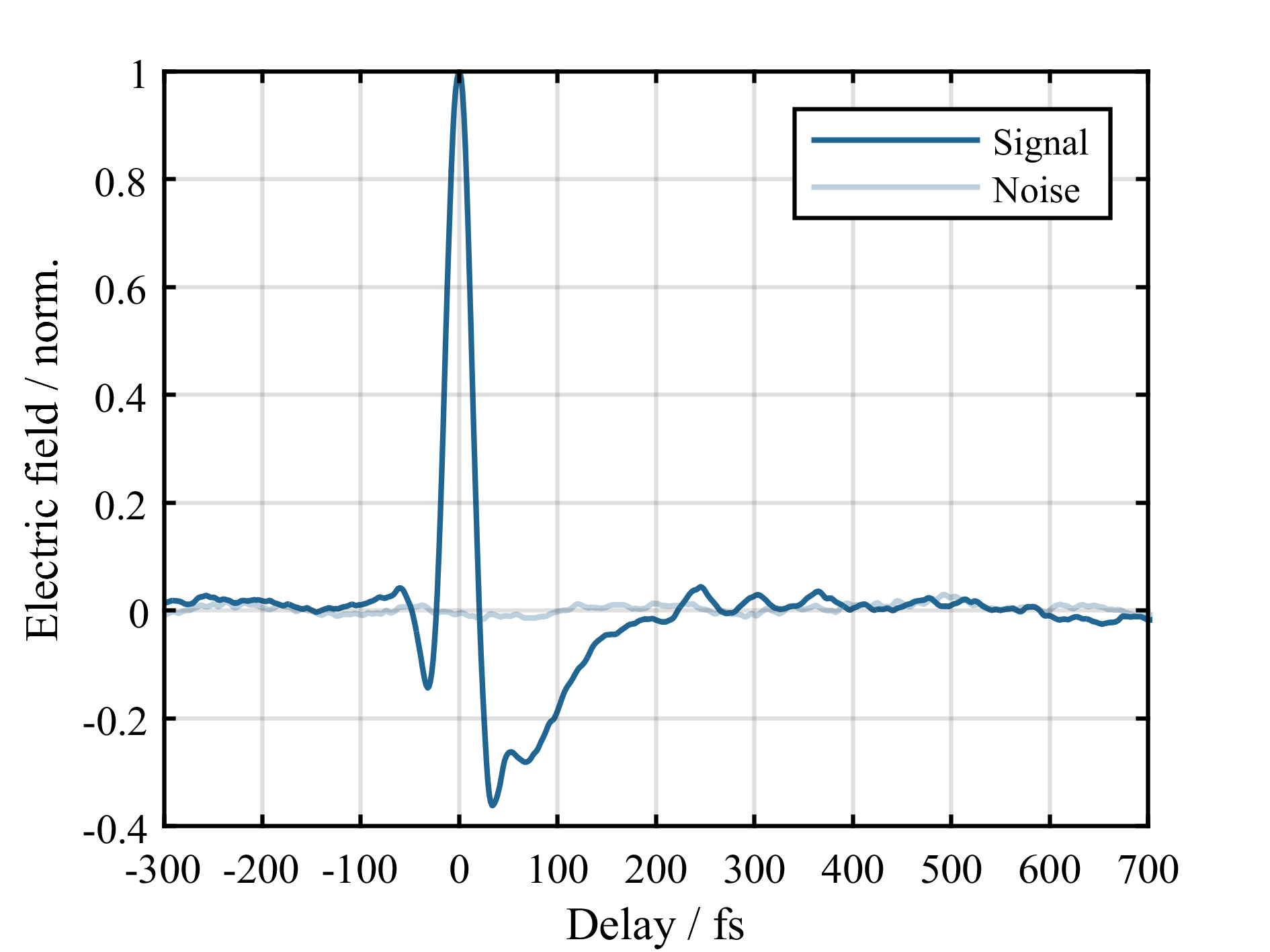}
\includegraphics[width=\linewidth]{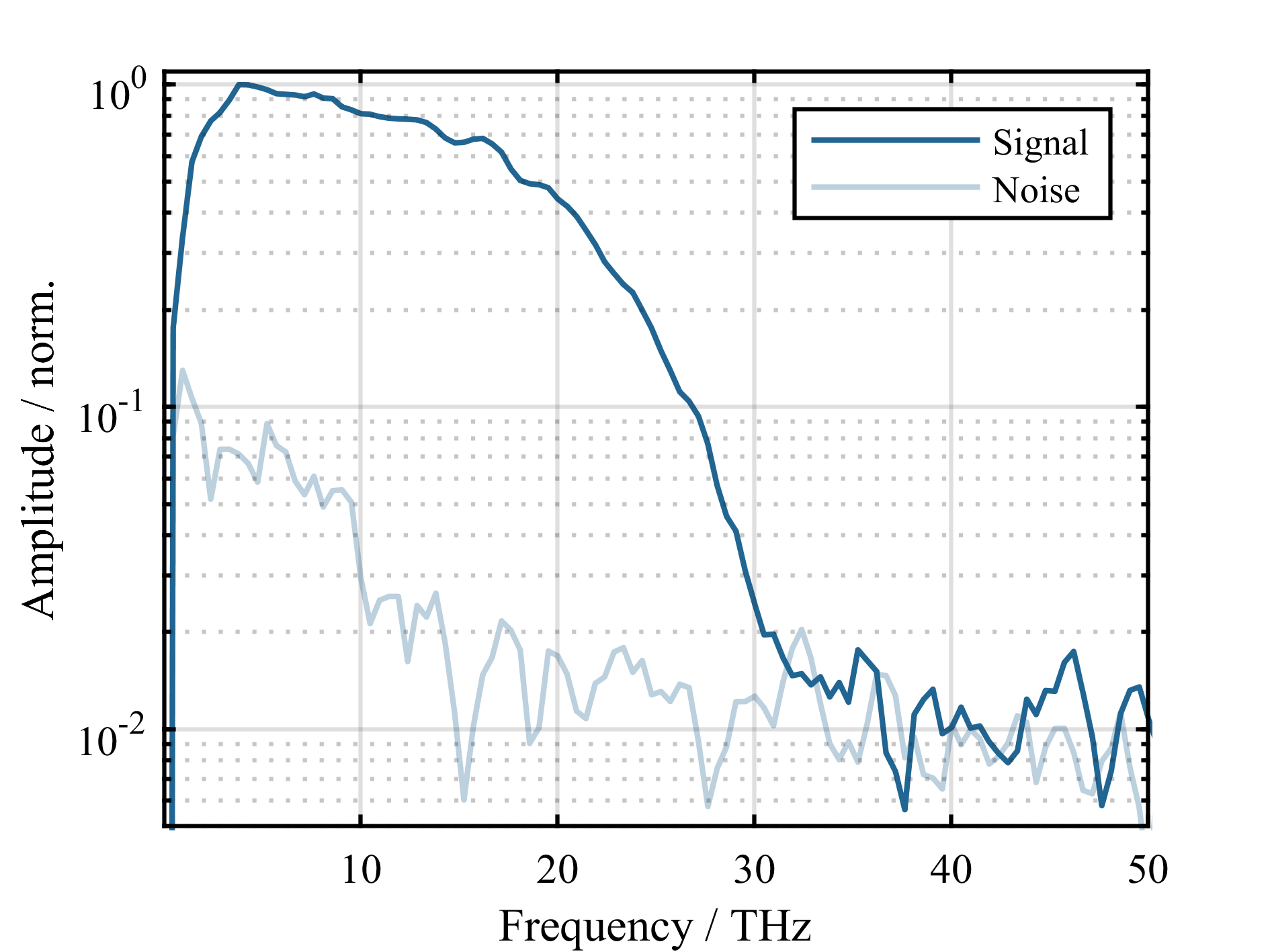}
\caption{(Top) Single-cycle THz pulse measured with the ABCD setup. (Bottom) Spectrum retrieved by Fourier-transforming the ABCD-trace.}
\label{fig:thz}
\end{figure}

\begin{figure}[!tb]
\centering
\setlength{\unitlength}{0.1\linewidth}
\begin{picture}(10,8)
\put(0,0){\includegraphics[width=\linewidth]{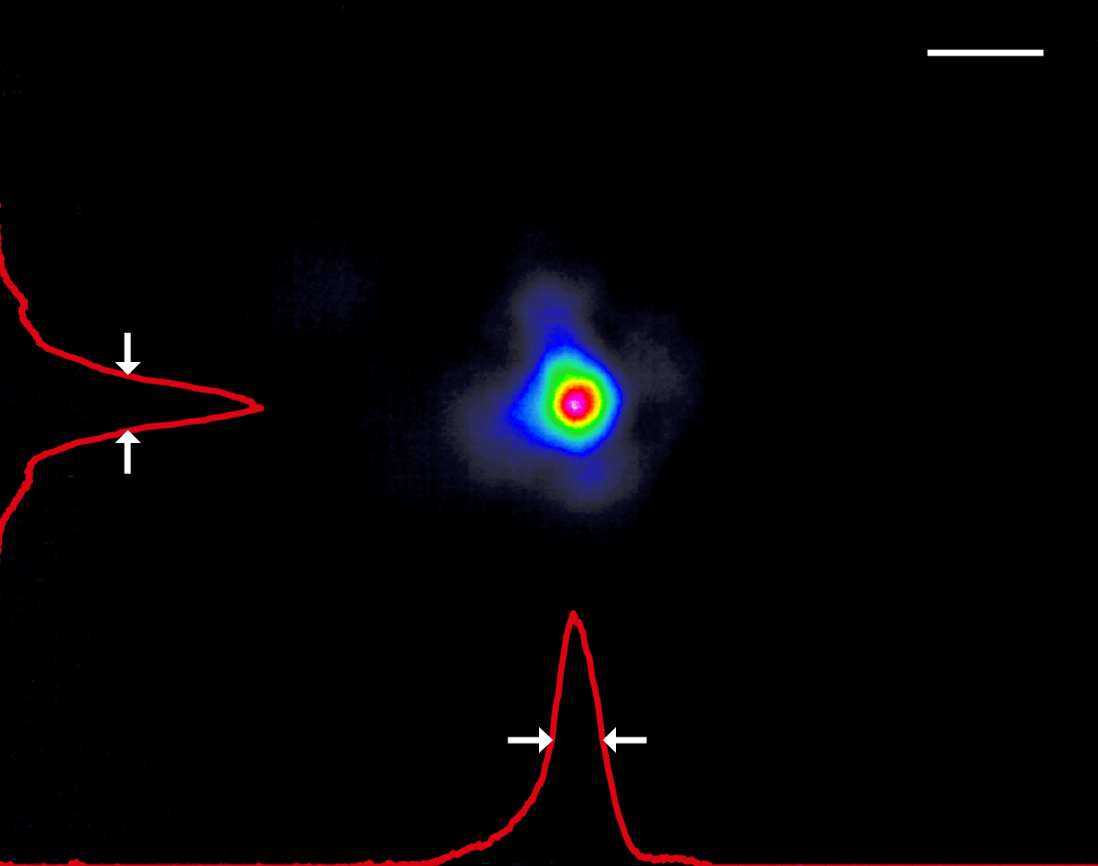}}
\put(8.55,7){\textcolor{white}{$ \SI{2}{\milli\meter}$}}
\put(6,1.01){\textcolor{white}{$ \SI{885}{\micro\meter}$}}
\put(1.07,3.2){\textcolor{white}{$ \SI{990}{\micro\meter}$}}
\end{picture}
\caption{Beam profile of the THz radiation focused with a 500 mm focal length gold mirror. Integrating over the beam reveals a power content of $ \SI{63}{\percent} $ inside the $ 1/e^2 $ beam diameter. The red curves show the beam profiles along the x- and y-axis with the arrows marking the FWHM diameter along each profile.}
\label{fig:focus}
\end{figure}
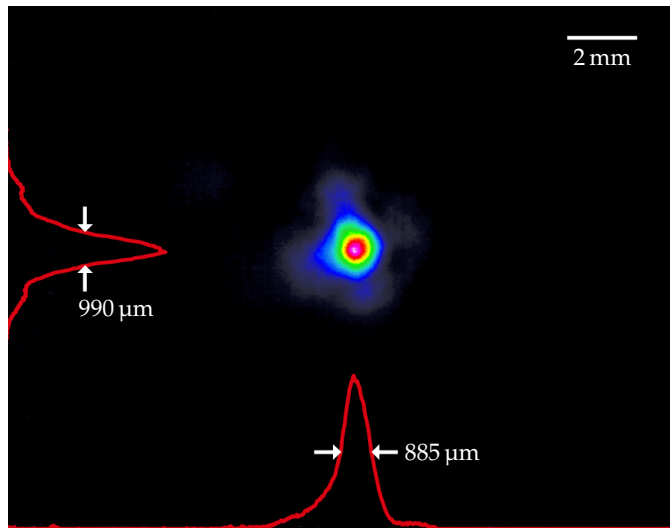

In the upper plot of Fig. \ref{fig:thz} the temporal THz electric field measured with the ABCD setup is depicted. The temporal shape shows a single-cycle feature. The corresponding spectrum retrieved by Fourier-transforming the temporal electric field covers the whole THz gap with spectral content up to 30 THz. The power of the generated THz radiation was measured to be $ \SI{640}{\milli\watt} $. Together with the average power of the driving pulses used for the generation process, which is 90 \% of $ \SI{633}{\watt} $, an efficiency of $ 1.1\cdot10^{-3} $ has been achieved. The pulse energy of the THz pulses is $ \SI{1.3}{\micro\joule} $. An image of the focal spot is shown in Fig. \ref{fig:focus}, indicating an energy content of $ 63 \% $ in the $ 1/e^2 $ spot diameter.
From the focal spot, the electric field measurement and the measured power of THz radiation it is possible to calculate the peak intensity and THz electric field achievable with this source. When focusing the beam with a $ \SI{50}{\milli\meter} $ focal length OAP a peak intensity in the order of $ \SI{10}{\giga\watt\per\centi\meter\squared} $ and a peak electric field in the order of $ \SI{1}{\mega\volt\per\centi\meter} $ can be achieved.
The red curves in Fig. \ref{fig:focus} show the profile along the x- and y-axis through the peak of the beam.

In conclusion, we have shown the generation of broadband THz radiation with an average power of $ \SI{640}{\milli\watt} $. This is the highest average power of laser-generated THz radiation achieved up to date. By implementing a gas-jet generation scheme, the efficiency of the generation process has been increased to over $ 1.1\cdot 10^{-3}$, exceeding previous experiments with an efficiency in the range of $ 10^{-4} $. The increase in efficiency has been allowed by the high pressure in the focus as well as the precise alignment and optimization enabled by the parallel and real-time measurement of the temporal electric field, the beam profile and the average THz power. The generated radiation covers the whole THz spectral region from $ \SI{0.1}{\tera\hertz} $ to $ \SI{30}{\tera\hertz} $, which enables broadband spectroscopy of even strongly absorbing samples. The strong single cycle feature of the pulse allows driving nonlinear effects and the studies of nonlinear material properties.
In the near future the modification of the setup to accommodate a Thulium-doped laser with $ \SI{2}{\micro\meter} $ driving wavelength is planned. 
With the wavelength scaling of the efficiency of the generation process reported in literature \cite{WavelengthScaling0,WavelengthScaling1,WavelengthScaling2} an increase of the efficiency by roughly one order of magnitude can be expected.
That will allow the generation of watt-class THz radiation from only $ \SI{100}{\watt} $ laser power, which paves the way to widely applicable, compact, broadband, watt-level THz sources.

\section*{Funding}
This work was supported by the European Research Council (ERC) under the European Union’s Horizon 2020 research and innovation programme (grant 835306, SALT), the Fraunhofer Cluster of Excellence Advanced Photon Sources (CAPS) and the Bundesministerium für Bildung und Forschung (13N15244, funding program Photonics Research Germany). C. J. acknowledges funding by the Deutsche Forschungsgemeinschaft (DFG, German Research Foundation), 416342637.

\section*{Disclosure}
The authors declare that there are no conflicts of interest related to this article.

\section*{Data availability}
Data underlying the results presented in this paper are not publicly available at this time but may be obtained from the authors upon reasonable request.

\bibliography{references}



\end{document}